\documentclass[a4paper,11pt]{article}
\usepackage{pos}

\usepackage{mciteplus}

\newcommand{\be}{\begin{equation}}
\newcommand{\ee}{\end{equation}}
\newcommand{\ri}{{\rm i}}
\newcommand{\la}{\langle}
\newcommand{\ra}{\rangle}

\title{Finite temperature and $\delta$-regime in the 2-flavor Schwinger model}

\author*[a]{Ivan Hip}
\author[b,c]{Jaime Fabi\'{a}n Nieto Castellanos}
\author[c]{Wolfgang Bietenholz}

\affiliation[a]{ Faculty of Geotechnical Engineering, University of Zagreb \\
  Hallerova aleja 7, 42000 Vara\v{z}din, Croatia}

\affiliation[b]{
  Facultad de Ciencias, Universidad Nacional Aut\'{o}noma de M\'{e}xico \\
  A.P.\ 70-542, C.P.\ 04510 Ciudad de M\'{e}xico, Mexico}

\affiliation[c]{
  Instituto de Ciencias Nucleares, Universidad Nacional Aut\'{o}noma de
  M\'{e}xico \\
A.P. 70-543, C.P. 04510 Ciudad de M\'{e}xico, Mexico}

\emailAdd{ivan.hip@gfv.unizg.hr}
\emailAdd{jafanica@ciencias.unam.mx}
\emailAdd{wolbi@nucleares.unam.mx}

\abstract{The Schwinger model is often used as a testbed for conceptual
and numerical approaches in lattice field theory. Still, some of its rich
physical properties in anisotropic volumes have not yet been explored.
For the multi-flavor finite temperature Schwinger model there is an
approximate solution by Hosotani {\it et al.}\ based on bosonization.
We perform lattice simulations and check the validity of this
approximation in the case of two flavors.
Next we exchange the r\^{o}le of the coordinates to enter the
$\delta$-regime, and measure the dependence of the residual ``pion''
mass on the spatial size, at zero temperature. Our results show that
universal features, which were derived by Leutwyler, Hasenfratz
and Niedermayer referring to quasi-spontaneous symmetry breaking
in $d>2$, extend even to $d=2$. This enables the computation of
the Schwinger model counterpart of the pion decay constant $F_{\pi}$.
It is consistent with an earlier determination by Harada {\it et al.}
who considered the divergence of the axial current in a light-cone
formulation, and with analytical results that we conjecture from 2d
versions of the Witten--Veneziano formula and the
Gell-Mann--Oakes--Renner relation, which suggest $F_{\pi} = 1/ \sqrt{2\pi}$.}

\FullConference{%
 The 38th International Symposium on Lattice Field Theory, LATTICE2021
  26th-30th July, 2021
  Zoom/Gather@Massachusetts Institute of Technology
}

\begin{document}
\maketitle

\section{The multi-flavor Schwinger model}

The Schwinger model represents Quantum Electrodynamics ---
fermions coupled to an Abelian gauge field --- in $d=2$ space-time
dimensions \cite{Schwinger1962a}. The original 1-flavor version was
solved analytically, which revealed in particular an axial anomaly,
whereas for $N_{\rm f} > 1$ flavors the chiral condensate
vanishes in the chiral limit. The multi-flavor version is still of
interest, as we see from a number of contributions to this conference.

The Schwinger model shares qualitative features with
QCD, in particular confinement and topology of the gauge field
configurations. It does not capture, however, asymptotic freedom (the
gauge coupling constant $g$ is indeed constant), nor spontaneous chiral
symmetry breaking. Still, the spectrum contains $N_{\rm f}-1$
light bosons; at finite fermion mass and/or in finite volume, their
behavior is similar to quasi-Nambu-Goldstone bosons. By analogy,
and in agreement with the literature, we denote them as ``pions''.

In addition, the particle spectrum contains a heavier boson, which can
be interpreted as the ``photon'', but --- following another analogy ---
it is often denoted as the ``$\eta$-meson''. Since it is a flavor singlet,
its 3-flavor QCD analogue is $\eta_{1}$, which is close to $\eta'$,
but in the Schwinger model we also just call it $\eta$.
In the chiral limit, its mass is given by \cite{Belvedere1979}
\be  \label{meta2}
m_{\eta}^{2} = \frac{N_{\rm f} g^{2}}{\pi}
\ee
(the coupling $g$ has mass dimension 1), while the pion is massless.
At finite (degenerate) fermion mass $m$, no exact solution is
known, but an approximate solution predicts the pion mass in infinite
volume as \cite{Hetrick1995}
\be  \label{mpi}
m_\pi = 4e^{2\gamma}\sqrt{\frac{2}{\pi}} \
      (m^2 g)^{1/3} = 2.1633 \dots (m^2 g)^{1/3} \ ,
\ee
where $\gamma = 0.577\dots$ is Euler's constant.

We present simulation results on regular, Euclidean lattices,
with Wilson fermions and the plaquette gauge action, obtained
with the Hybrid Monte Carlo algorithm. We are particularly
interested in anisotropic volumes: first, we study
this model at finite temperature, and compare the ``meson'' masses
with theoretical predictions. A bosonization ansatz reduces the system
to a quantum mechanical problem \cite{Hetrick1995,Hetrick1996}, which
we solve numerically.

By inverting the r\^{o}le of the coordinates, we access the
$\delta$-regime, which is still unexplored in $d=2$. We conjecture,
and confirm, a residual pion mass $m_{\pi}^{\rm R} \propto 1/L$ at
$m=0$. The proportionality constant provides a value for a
parameter, which we denote --- by analogy --- as the ``pion decay
constant'' $F_{\pi}$. It is dimensionless in $d=2$, and its value
is consistent with the Witten-Veneziano relation (if we identify
$F_{\pi} = F_{\eta}$), and with the
Gell-Mann--Oakes--Renner relation. It further agrees with a previous
determination in the framework of a light-cone formulation, which refers
to the divergence of the axial current \cite{Harada:1993va}.

\section{The masses $m_{\pi}$ and $m_{\eta}$ at finite temperature}

\begin{figure}[h!]
  \begin{center}
 \includegraphics[width=0.5\textwidth]{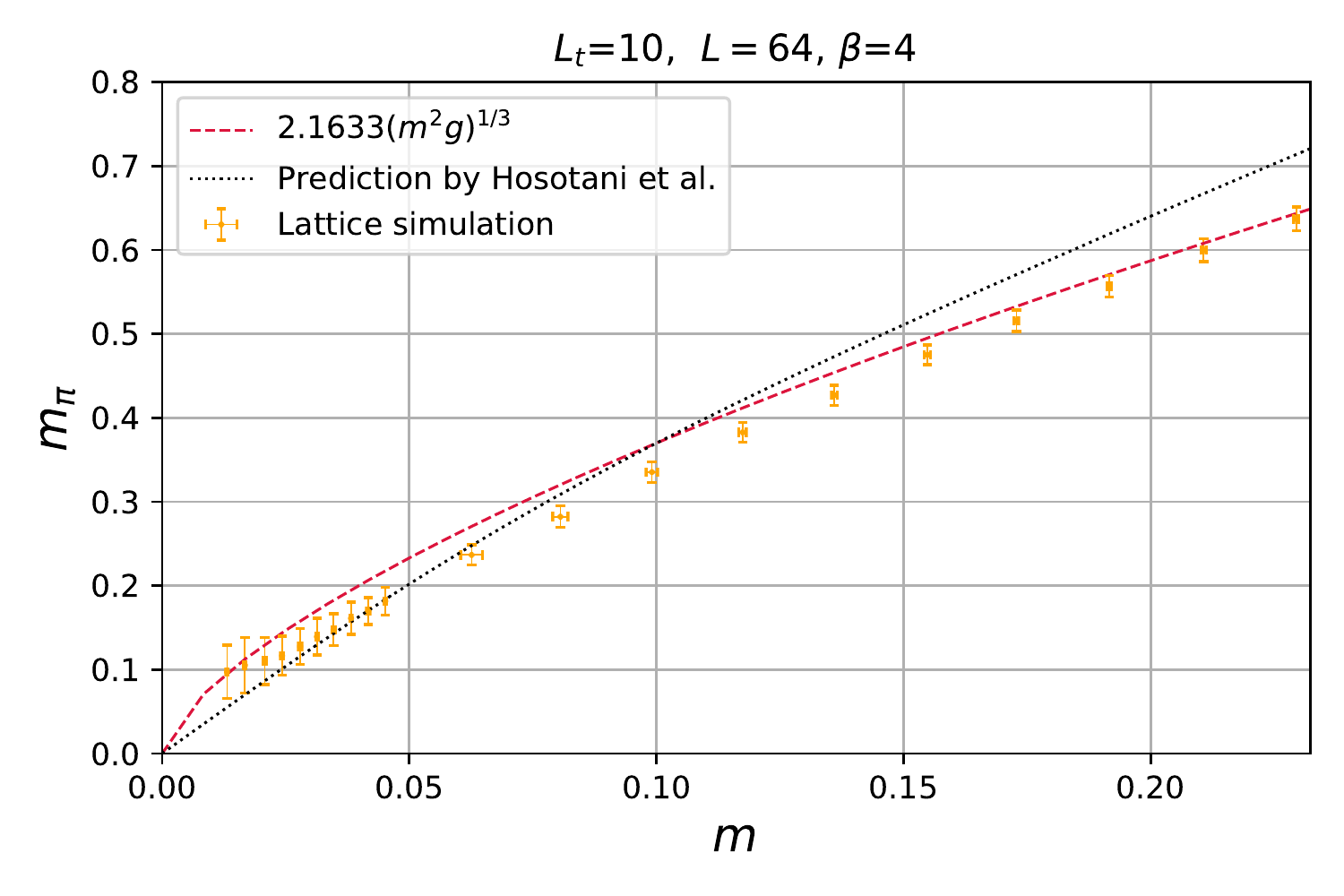}
    \hspace*{-2mm}
    \includegraphics[width=0.5\textwidth]{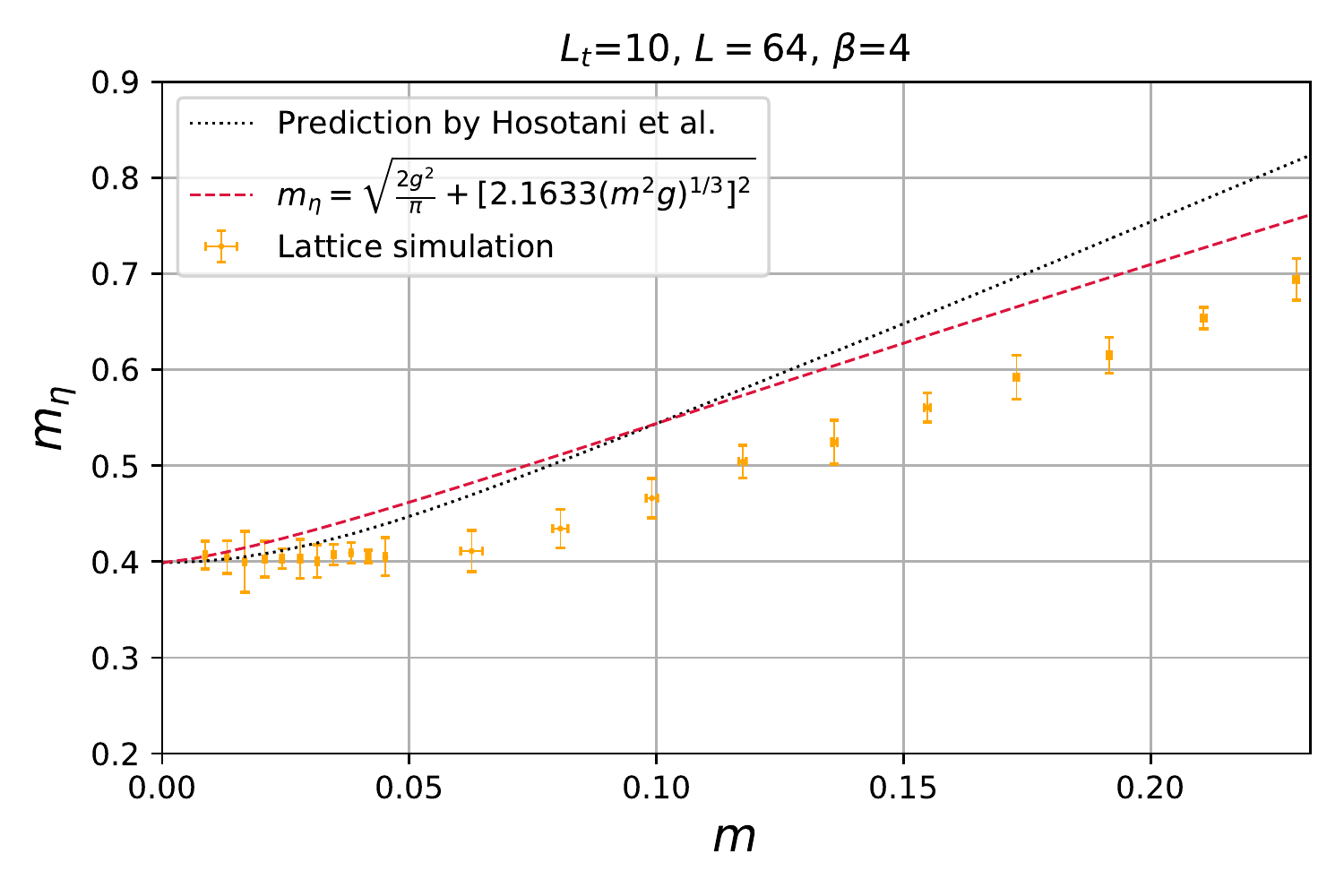}
  \end{center}
  \vspace*{-8mm}
  \caption{The masses $m_{\pi}$ and $m_{\eta}$ as functions of fermion
    mass $m$, at finite temperature, obtained from the approximation
    (\ref{mpi}) of Ref.\ \cite{Hetrick1995}, from a bosonization method
    \cite{Hosotani95, *Hosotani98}, and from lattice simulations.}
 \vspace*{-2mm}
  \label{mpifiniteT}
\end{figure}

In the 1990s, Hetrick, Hosotani and Iso discussed
the bosonization of the multi-flavor Schwinger model
\cite{Hetrick1995,Hetrick1996}. Here we particularly refer to a
system of non-linear differential equations given in Refs.\
\cite{Hosotani95, *Hosotani98},
which represent a quantum mechanical description of the model at
finite temperature. These are Schr\"{o}dinger-type equations,
which imply the values of $m_{\pi}$, $m_{\eta}$ and the chiral
condensate $\Sigma$, as functions of the degenerate fermion mass $m$.
We solved them numerically, as an eigenvalue problem, by three
numerical methods. They lead to consistent results, which stabilize
for increasing matrix size.
Figure \ref{mpifiniteT} shows these results for $m_{\pi}$ and $m_{\eta}$,
as functions of $m$, for
$N_{\rm f}=2$ flavors.\footnote{For the chiral condensate, obtained
  from bosonization, we refer to Ref.\ \cite{Jaime}.}
The pion mass is compared to the approximation of eq.\ (\ref{mpi}),
which predicts a larger (smaller) $m_{\pi}$ at small (moderate) $m$.

As a test, we measured $m_{\pi}$ and $m_{\eta}$ by simulations
on a $L_{t} \times L = 10 \times 64$ lattice.
In this case, the (renormalized) fermion mass
is measured by referring to the PCAC relation. Simulations at various
values of $\beta \equiv 1/g^{2}$ show that the lattice artifacts are small
at $\beta =4$.\footnote{We are using lattice units. For a general lattice
  spacing $a$, this relation takes the form $\beta \equiv 1/(a g^{2})$.}
The notorious problems close to the chiral limit prevent reliable results at
$m \lesssim 0.02$.
In the range of $0.02 \lesssim m \lesssim 0.05$ the bosonization prediction
is compatible with the simulation results. At larger fermion mass, however,
this approximation significantly overestimates both $m_{\pi}$ and $m_{\eta}$.
On the other hand, around $m \approx 0.2$ formula (\ref{mpi}) for
$m_{\pi}$ is in agreement with the simulation results.

The formulae of Refs.\ \cite{Hosotani95, *Hosotani98} also
include the case of an arbitrary vacuum angle $\theta$, which could
be of interest to probe simulation methods which try to overcome the
sign problem. However, here we see that these formulae are only
reliable at small $m$, where the simulations are confronted with
additional difficulties.

\section{\texorpdfstring{Residual pion mass in the $\delta$}{d}-regime}

Chiral perturbation theory, as a systematic effective low-energy theory
for QCD,  distinguishes the regimes of large space-time volume ($p$-regime),
small space-time volume ($\epsilon$-regime) and small spatial volume
but a large extent in (Euclidean) time, $L \ll L_{t}$ ($\delta$-regime);
the length scale is set by the inverse pion mass.

Here we address the $\delta$-regime, which is least explored, and where
finite-size effects entail a residual pion mass $m_{\pi}^{\rm R}$ even in
the chiral limit.
It was introduced by Leutwyler \cite{Leutwyler1987}, who approximated
the quasi-1d system by quantum mechanics, such that $m_{\pi}^{\rm R}$
corresponds to the mass gap of a quantum rotor,
\be
m_{\pi}^{\rm R} = \frac{N_{\pi}}{2\Theta} \ , \quad
\Theta \simeq F_{\pi}^{2} L^{3} \ .
\ee
$N_{\pi}$ is the number of pions, and $\Theta$ is the moment of inertia,
which is given here to leading order, in $d=4$ \cite{Leutwyler1987}.

In the framework of O($N$) models, with $N_{\pi} = N-1$,
Hasenfratz and Niedermayer generalized
this formula with respect to the space-time dimension $d>2$,
and computed $\Theta$ to next-to-leading order \cite{Hasenfratz1993},
\be
\Theta = F_\pi^2 L^{d-1} \left[1 +
          \frac{N_{\pi} - 1}{2\pi F_\pi^2 L^{d-2}}
          \left(\frac{d - 1}{d - 2} + \dots \right) \right] \ .
\ee
We see that $F_{\pi}$ has the mass dimension $d/2-1$.
The restriction to $d>2$ avoids a possible singularity in the last term,
in agreement with the concept of would-be Nambu-Goldstone bosons in
infinite volume.

There are only few lattice QCD studies in the $\delta$-regime.
In the next-to-next-to-leading order, sub-leading low-energy constants
appear \cite{Hasenfratz10, *Niedermayer16}, and the comparison with QCD
simulation results led in particular to a reasonable value of the
controversial constant $l_{3}$ \cite{Bietenholz2010}.
The transition to the $p-$ and $\epsilon$-regime is investigated
in Ref.\ \cite{Matzelle16}.

\begin{figure}[h!]
\hspace*{-3mm}
\includegraphics[width=0.55\textwidth]{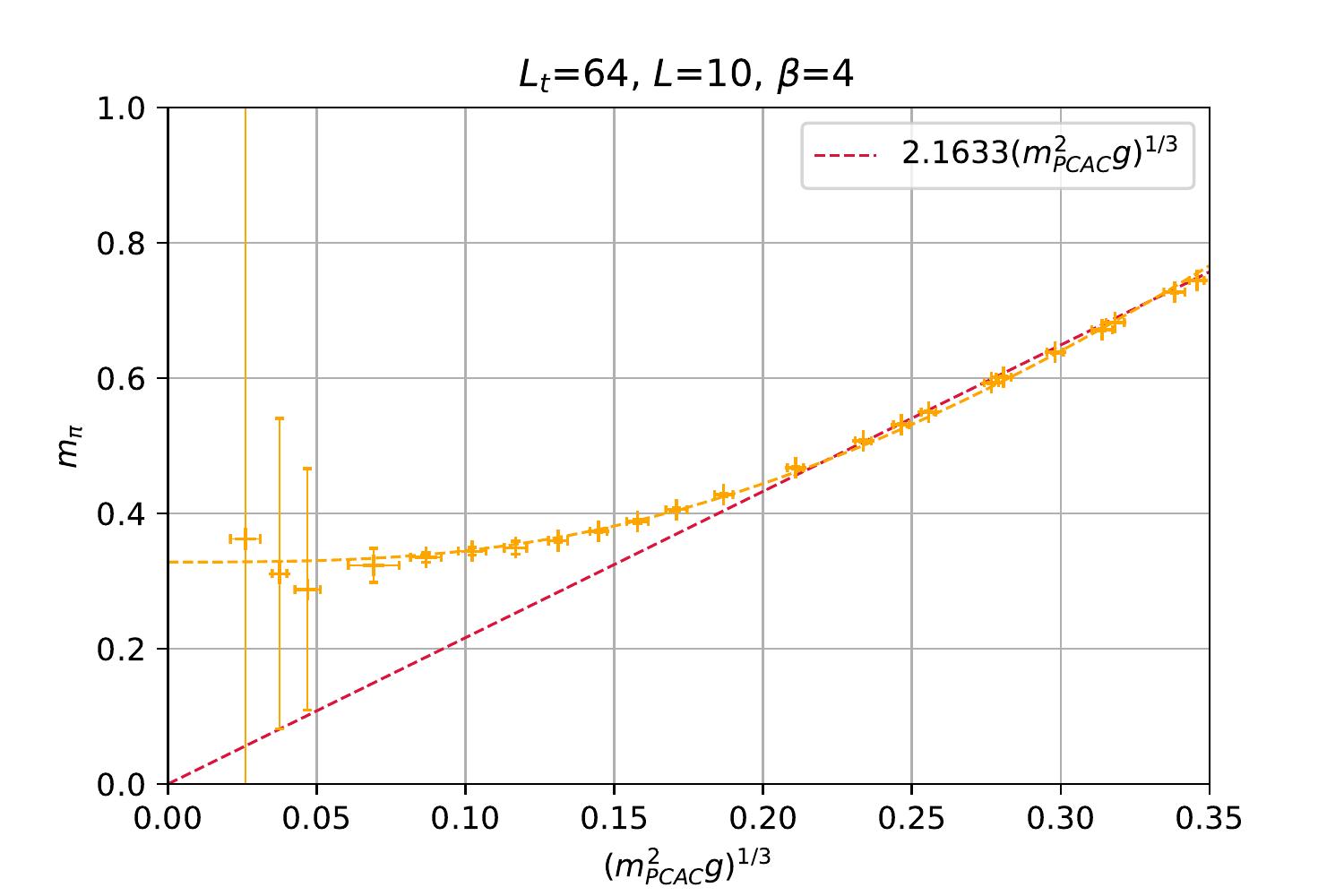}
\hspace*{-7mm}
\includegraphics[width=0.55\textwidth]{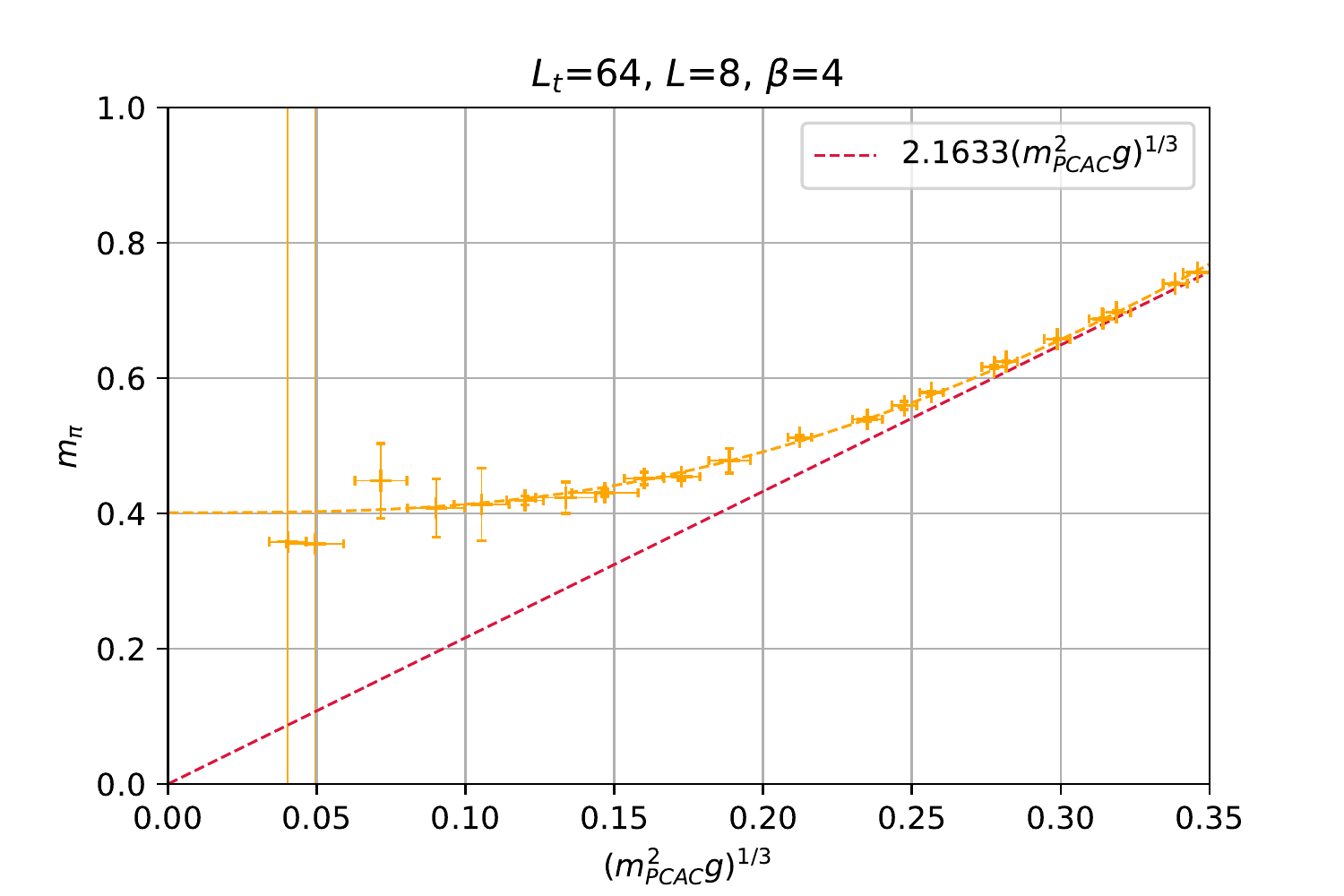}
  \caption{Illustration of the residual pion mass plateaux
    in spatial sizes $L =10$ and $L=8$, at $\beta = 4$.}
\label{MpiRextra}
\end{figure}

In our case, the next-to-leading order term has the prefactor $N_{\pi}-1=0$.
We dismiss it, despite the denominator $d-2$, so we conjecture for
the 2-flavor Schwinger model
\be  \label{MpiRformula}
m_{\pi}^{\rm R} \simeq \frac{1}{2F_{\pi}^{2} L} \ .
\ee
In order to test this conjecture, we performed simulations on
lattices with spatial size $L \ll L_{t} =64$, at $\beta =3$, $4$ and
$5$. The value of $m_{\pi}^{\rm R}$ is obtained by a chirally extrapolated
plateau; two examples are illustrated in Figure \ref{MpiRextra}.

\begin{figure}[h!]
  \vspace*{-2mm}
  \begin{center}
    \hspace*{-2mm}
    \includegraphics[width=0.53\textwidth]{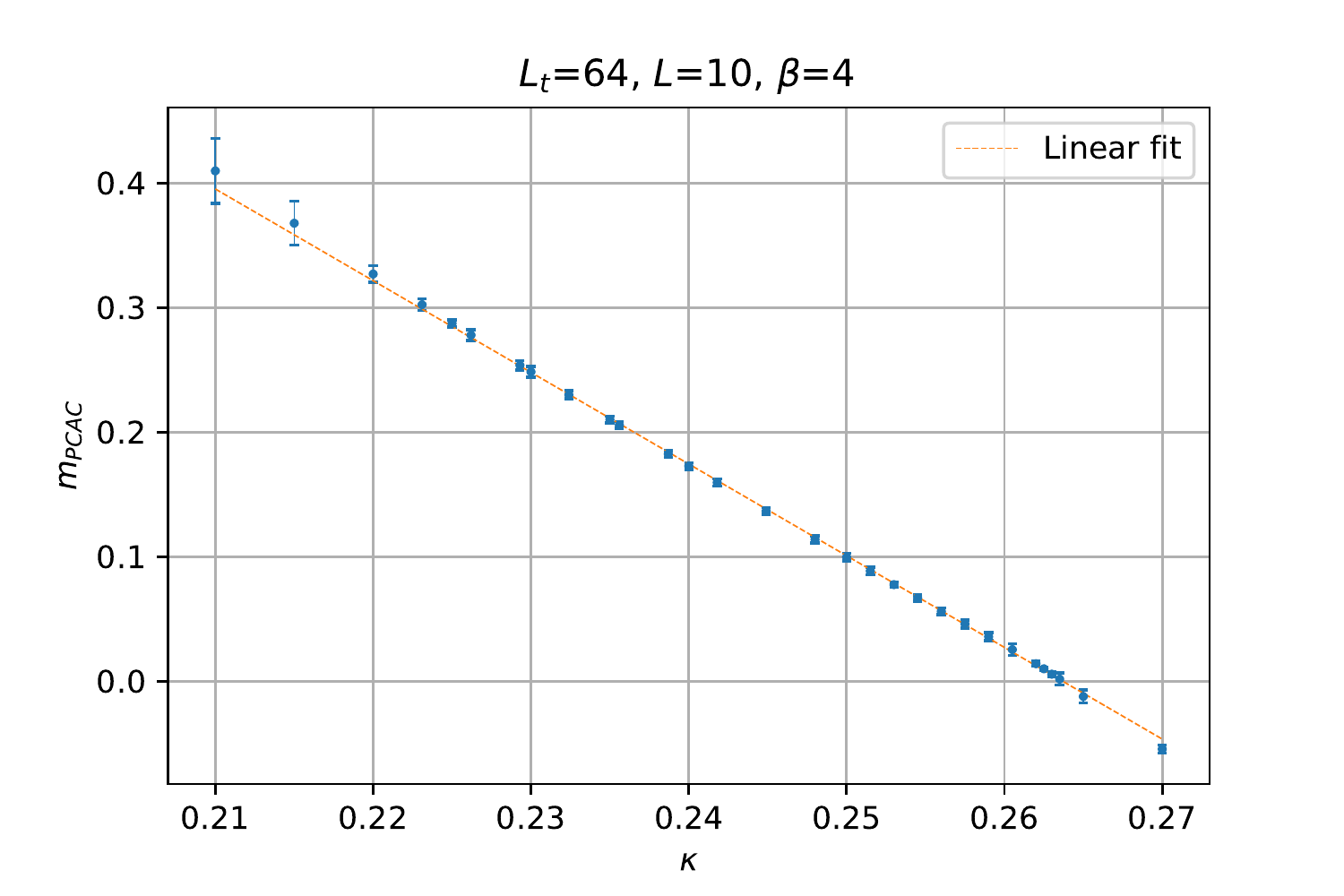}
    \hspace*{-9mm}
    \includegraphics[width=0.53\textwidth]{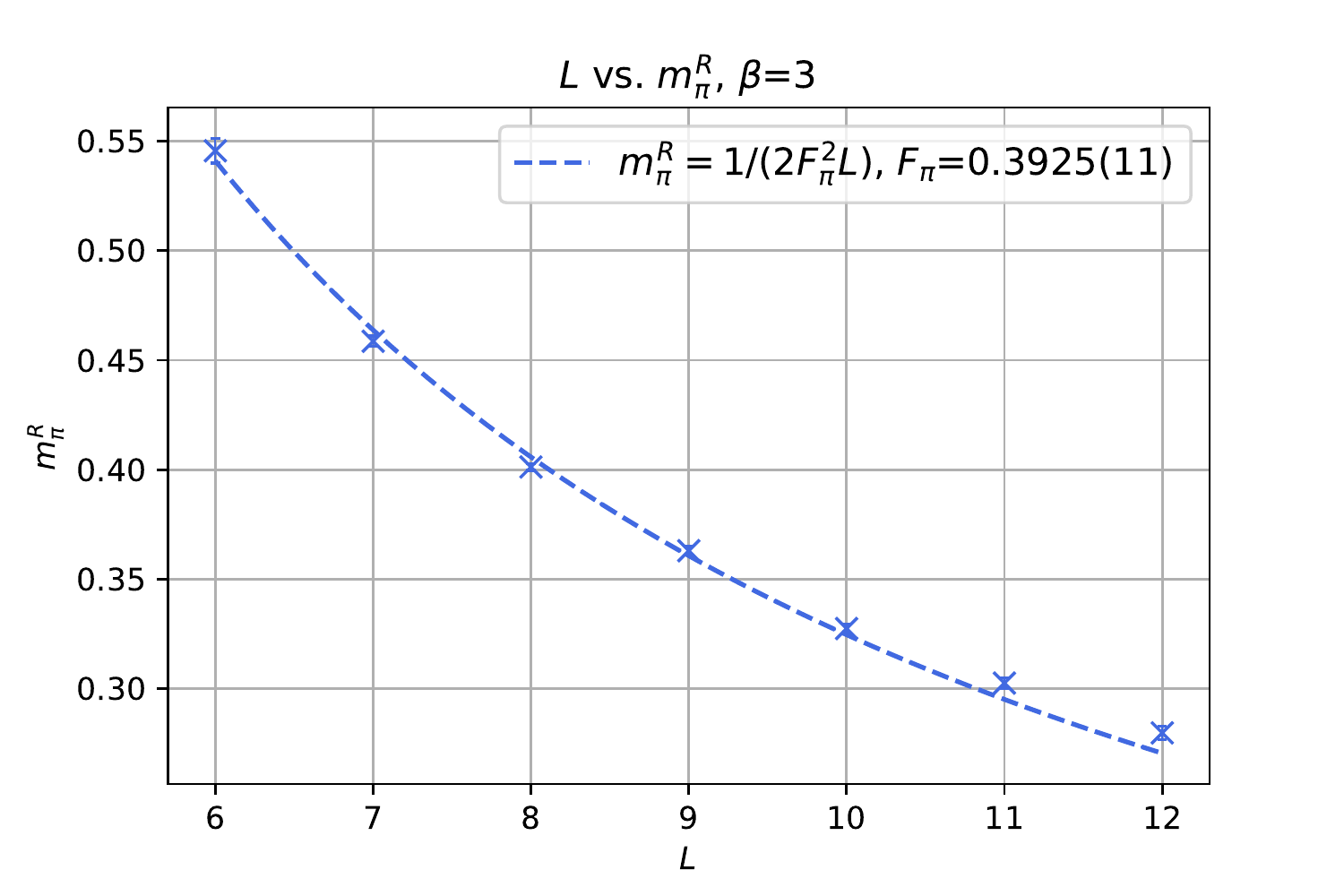} \hspace*{-2mm} \\
    \hspace*{-2mm}
    \includegraphics[width=0.53\textwidth]{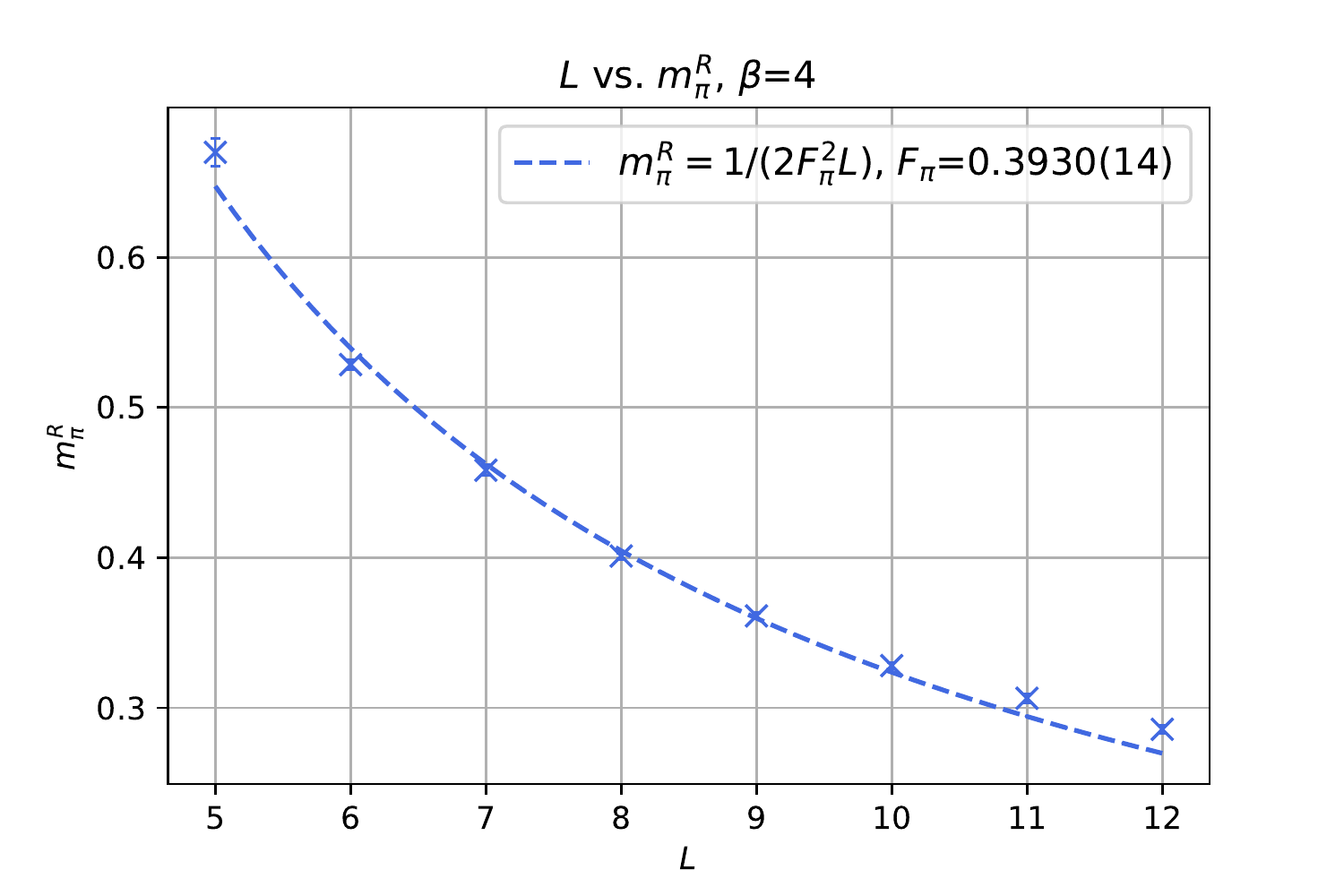}
    \hspace*{-9mm} 
    \includegraphics[width=0.53\textwidth]{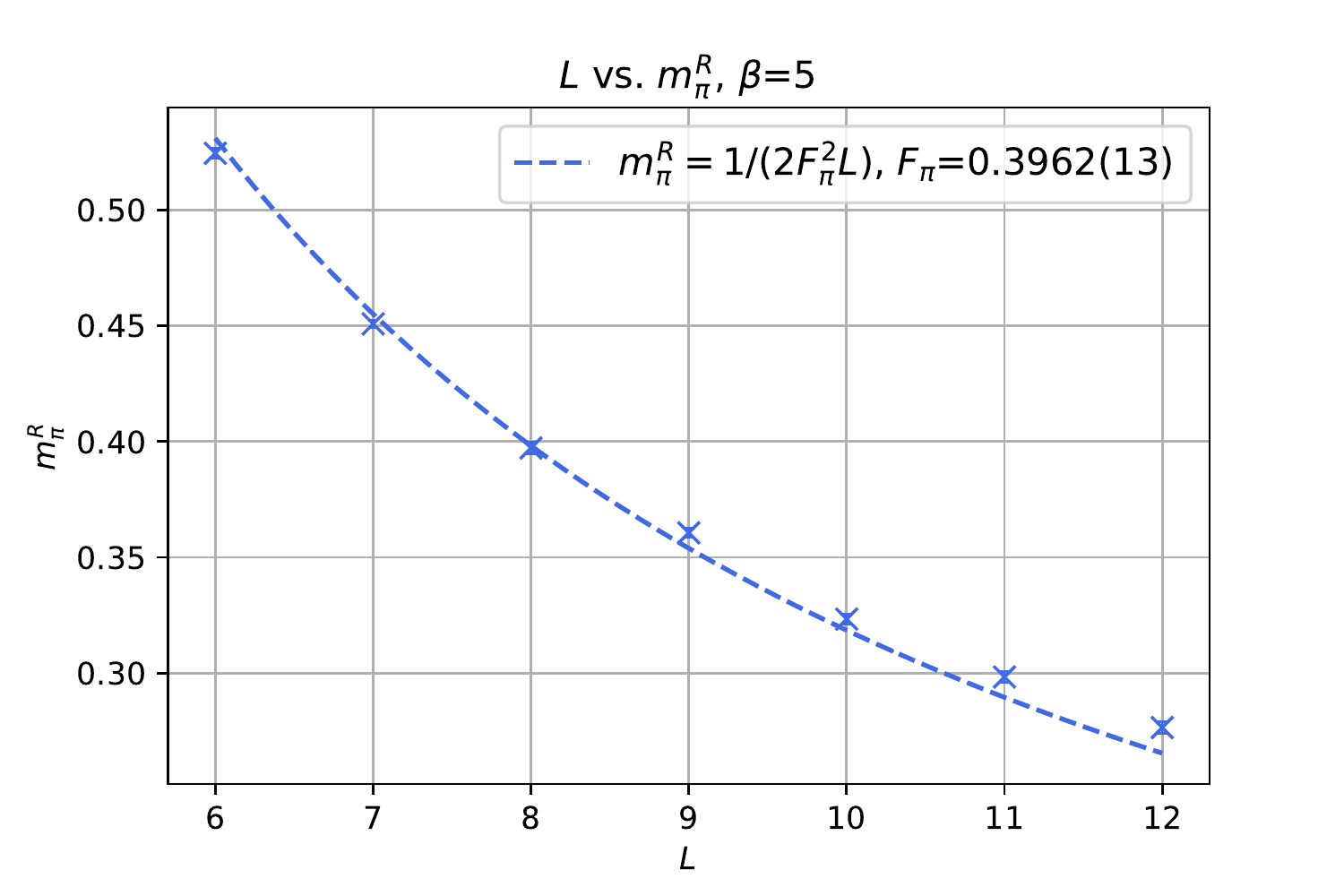} \hspace*{-2mm}
  \end{center}
  \vspace*{-4mm}
  \caption{Top, left: Fermion mass $m_{\rm PCAC}$ as a function of the hopping
    parameter $\kappa$, in one example. Generally we observe an approximately
    linear behavior. Rest:
    Residual pion mass $m_{\pi}^{\rm R}$ in the $\delta$-regime,
    as a function of the spatial size $L$, at fixed $\beta \equiv 1/g^{2}$.
    The simulation results are consistent with the hypothesis
    $m_{\pi}^{\rm R} \propto 1/L$. In each case, a 1-parameter fit to
    relation (\ref{MpiRformula}) provides
    the value of $F_{\pi}$ in Table \ref{tabFpidelta}.}
\label{MpiRvsL}
\end{figure}

The plots in Figure \ref{MpiRvsL} show an example for the PCAC fermion
mass depending on the hopping parameter $\kappa$, and a multitude of
results for $m_{\pi}^{\rm R}$ at fixed $\beta$, but different $L$.
We observe good agreement with the
conjectured proportionality $m_{\pi}^{\rm R} \propto 1/L$.

This property allows us to proceed and extract the ``pion decay
constant'' according to eq.\ (\ref{MpiRformula}). The fits at fixed
$\beta$ lead to the $F_{\pi}$-values in Table \ref{tabFpidelta}.
\begin{table}[h!]
 \begin{center}
  \begin{tabular}{|c||c|c|c|}
        \hline   
	$\beta \equiv 1/g^{2} $ & 3 & 4 & 5 \\
        \hline
        $F_\pi$  & 0.3925(11) & 0.3930(14) & 0.3962(13) \\
	\hline
  \end{tabular}
 \end{center}
 \vspace*{-2mm}
 \caption{Results for $F_{\pi}$, obtained by fits to eq.\ (\ref{MpiRformula}),
   at three values of $\beta$.}
 \label{tabFpidelta}
 \vspace*{-2mm}
\end{table}
The results at $\beta =3$, $4$ and $5$ agree to percent level, but
for increasing $\beta$ ({\it i.e.}\ suppressed lattice artifacts)
we observe a slight trend up --- we will come back to it.

\section{The 2d Witten--Veneziano formula}

In the large-$N_{\rm c}$ limit of QCD, at finite 't Hooft coupling
$g_{\rm s} \sqrt{N_{\rm c}}$, the 3-flavor chiral symmetry breaking has the
structure ${\rm U}(3) \otimes {\rm U}(3) \to {\rm U}(3)$. This implies
9 Nambu-Goldstone bosons, which include --- in addition to the meson
octet built of $\pi$, $K$ and $\eta$ --- the $\eta'$-meson.
If one considers $1/N_{\rm c}$-corrections, the latter picks up a mass,
which (with massless quarks $u$, $d$, $s$) is given by
the Witten--Veneziano formula
\cite{Witten1979, *Veneziano1979},
$ m_{\eta'}^{2} F_{\eta'}^{2} = 2 N_{\rm f} \chi_{\rm t}^{\rm q} $,
where $\chi_{\rm t}^{\rm q}$ is the quenched topological susceptibility:
to this order, quark loops do not contribute, and $F_{\eta'} = F_{\pi}$. 
According to lattice simulation results for $\chi_{\rm t}^{\rm q}$, the
fact that the $\eta'$-meson is so heavy in Nature (heavier than a nucleon,
and therefore not interpretable as a quasi-Nambu-Goldstone boson) can
indeed be understood along these lines, as a topological effect.

According to Ref.\ \cite{Gattringer1994}, the conceptual basis of the
Witten--Veneziano relation is more solid in the multi-flavor Schwinger
model. In the chiral limit it reads
\be  \label{WV2d}
m_{\eta}^{2} = \frac{2N_{\rm f}}{F_{\eta}^{2}} \chi_{\rm t}^{\rm q} \ .
\ee
In this case there is no need to speculate (in QCD one assumes
$N_{\rm c}=3$ to behave similarly to large $N_{\rm c}$). On the other
hand, we do not have any obvious justification for setting
$F_{\pi} = F_{\eta}$, but we are going to consider this scenario
nevertheless.

Ref.\ \cite{Seiler1987} computed the topological susceptibility
in 2d U(1) pure gauge theory in the continuum, and infinite volume,
\be  \label{chitq} \qquad \qquad
\beta \chi_{\rm t}^{\rm q} = \beta \ ^{~~\lim}_{V \to \infty}
\, \frac{\la Q^{2} \ra}{V} = \frac{1}{4\pi^2} \qquad
(Q~:~{\rm topological~charge}) \ .
\ee
This value is consistent with the continuum limit of lattice
results for $\chi_{\rm t}^{\rm q}$. In particular, Ref.\ \cite{Bardeen1998}
considered the (non-integer) topological lattice charge
$Q_{\rm S} = \frac{1}{2\pi}\sum_{P}\sin(\theta_P)$, where $\theta_{P}$ is the
plaquette variable, and derived the exact expression
$\beta \chi_{\rm t}^{\rm q} = I_1(\beta) / [4 \pi^2 I_0(\beta)]$.

\begin{figure}
\vspace*{-4mm}
\begin{center}    
  \includegraphics[width=0.57\textwidth]{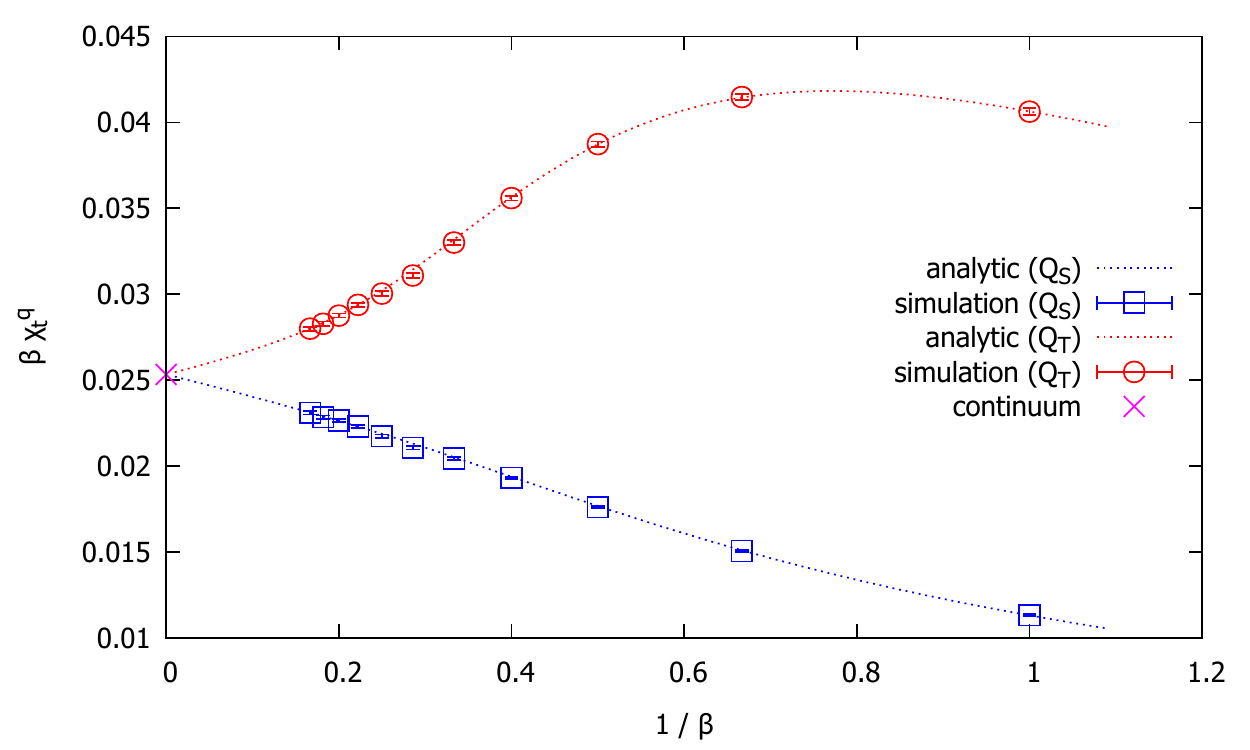}
\end{center}
\vspace*{-6mm}
\caption{The quenched topological susceptibility, for two different
  lattice formulations of the topological charge, based on the numerical
  evaluation of implicit expressions, and on simulations.}
\vspace*{-3mm}
\label{topsusquenched}
\end{figure}

If we refer to the usual lattice definition
$Q_{\rm T} = \frac{1}{2\pi}\sum_{P} \theta_P \in \mathbb{Z}$,
there is no closed expression for $\beta \chi_{\rm t}^{\rm q}$, but it
can be evaluated numerically to arbitrary precision \cite{Bonati19}.
Figure \ref{topsusquenched} shows both analytic expressions as functions
of $1/\beta$. As a consistency check we compare them to simulation results,
which accurately agree, and we also see that the continuum limit
coincides in both cases with eq.\ (\ref{chitq}).

Thus eq.\ (\ref{chitq}) is confirmed, and along with
eqs.\ (\ref{meta2}) and (\ref{WV2d}) we obtain, in the chiral limit,
\be
F_{\eta}^{2} = \frac{2N_{\rm f}}{m_{\eta}^{2}} \chi_{\rm t}^{\rm q}
= 2N_{\rm f} \ \frac{\pi \beta}{N_{\rm f}} \
\frac{1}{4\pi^{2} \beta} = \frac{1}{2\pi} \ ,
\ee
If we assume $F_{\pi} = F_{\eta}$, as in large-$N_{\rm c}$ QCD,
we obtain $F_{\pi} = 1 / \sqrt{2\pi} = 0.3989 \dots$, which is close
to the value of $F_{\pi}$ that we obtained in the $\delta$-regime,
given in Table \ref{tabFpidelta} --- in particular the continuum limit
seems perfectly compatible.

Finally we are now going to amplify our perspective and consider $F_{\pi}$
in the 2-flavor Schwinger model obtained by various formulations.

\section{The ``pion decay constant'' in the Schwinger model: an overview}

In QCD, the pion decay constant $F_{\pi}$ appears in a variety of
relations, for instance \vspace*{2mm} \\
(a) \hspace*{1mm} $ \la 0| J_{\mu}^{5}(0) | \pi (p) \ra = \ri p_{\mu} F_{\pi}$ \\
(b) \hspace*{1mm} $ \la 0| \partial_{\mu} J_{\mu}^{5}(0) | \pi (p) \ra = m_{\pi}^{2} F_{\pi}$ \\
(c) \hspace*{1mm} Coefficient to the leading term of $m_{\pi}^{\rm R}(L)$
in the $\delta$-regime \\
(d) \hspace*{1mm} Witten--Veneziano formula \\
(e) \hspace*{1mm} Gell-Mann--Oakes--Renner relation. \\

This list is incomplete, of course, one might add {\it e.g.}\ the
r\^{o}le as a leading Low Energy Constant in the pion effective
Lagrangian, the Goldstone-Wilczek current in the effective low-energy
theory for the neutral pion decay, or the coefficient of the
axial current correlation function in the $\epsilon$-regime (for
lattice studies, see {\it e.g.}\ Refs.\ \cite{Bietenholz04, *Giusti04}),
but in the following we are only going to refer to the relations (a) to (e).

Here the meaning of $F_{\pi}$ is always the same,
but this is not obvious anymore when we refer to one of these
relations to introduce --- by analogy --- a ``pion decay constant''
in the 2-flavor Schwinger model (although that ``pion'' does not decay).
To the best of our knowledge, the only previous study of this kind
was performed in Ref.\ \cite{Harada:1993va}, which referred to
relation (b). Working with a light-cone formulation
(at $m>0$), Harada, Sugihara and Taniguchi obtained
\be
F_{\pi}(m) = 0.394518(4) + 0.040(1) m / g \ .
\ee

In Section 3 we referred to property (c), and from the fits to
$m_{\pi}^{\rm R}(L)$ we obtained the values in Table \ref{tabFpidelta},
which agree to two digits.
Section 4 refers to relation (d), and if we add the hypothesis
$F_{\pi} = F_{\eta}$, we arrive at $F_{\pi} = 1/ \sqrt{2\pi}$.

Let us finally consider (e), the Gell-Mann--Oakes--Renner relation
in the Schwinger model \cite{Smilga92}
\be  \label{GMOR}
F_{\pi}^{2}(m) = \frac{2 m \Sigma}{m_{\pi}^{2}} \ ,
\ee
where $\Sigma = - \la \bar \psi \psi \ra$ is the chiral condensate.
Ref.\ \cite{Hetrick1995} derives explicit small-$m$ formulae for
$\Sigma$ and $m_{\pi}$. In a large volume (and at
vacuum angle $\theta =0$), the latter is consistent with
eqs.\ (\ref{meta2}) and (\ref{mpi}).
Inserting both into the Gell-Mann--Oakes--Renner relation (\ref{GMOR})
exactly confirms the result that we conjectured in Section 4, 
\be  \label{SigmaMpi}
\Sigma = \frac{1}{\pi} \Big( \frac{e^{4 \gamma} m \, m_{\eta}^{2}}{4}
\Big)^{1/3} \ ,
\quad m_{\pi} = \Big( 4 e^{2 \gamma} m^{2} m_{\eta} \Big)^{1/3} \quad
\Rightarrow \quad F_{\pi} = \frac{1}{\sqrt{2\pi}} \ . \quad
\ee
In this form, $F_{\pi}$ does not depend on $m$, nor on $m_{\eta}$,
and therefore neither on the coupling $g$.

Actually Ref.\ \cite{Hetrick1995} distinguishes (in its eqs.\ (36)
and (38)) three different regimes, depending on mass and size. In eq.\
(\ref{SigmaMpi}) we reproduced the formula for $\Sigma$ and $m_{\pi}$
which are valid if $m \sqrt{m_{\eta}} L^{3/2} \gg 1$, $m_{\pi} L \gg 1$
and $m_{\eta} \gg m_{\pi}$. Interestingly, when we insert in eq.\
(\ref{GMOR}) the formulae in any of the two other regimes,
the result for $F_{\pi}$ is exactly the same.\\

We conclude that relations (b), (c), (d) and (e) all lead to
values for the ``pion decay constant'' which are consistent with
$F_{\pi} = 1/\sqrt{2\pi}$, which looks highly satisfactory.\\

We close with two open questions:
\begin{itemize}

\item The consideration in Section 4, which refers to property (d),
  suggests the relation $F_{\pi} = F_{\eta}$ in the chiral limit.
  In fact, Ref.\ \cite{Gattringer1994} also predicts
  $F_{\eta} = 1/ \sqrt{2\pi}$ in the chiral limit of the 2-flavor
  Schwinger model, but its relation to $F_{\pi}$ remains to be understood.
  
\item Relation (a) is often considered the standard way to define
  $F_{\pi}$ in QCD. If we try to employ its analogue to define $F_{\pi}$
  in the Schwinger model, it seems to imply
  $F_{\pi}(m=0) = 0$,\footnote{We thank Stephan D\"{u}rr for pointing
    this out to us.}
  since the pions are sterile, {\it i.e.} free, if we are strictly in
  the chiral limit (this is how a contradiction with the
  Mermin-Wagner-Hohenberg-Coleman theorem is evaded \cite{SmiVer}).
  In light of the results presented here, also that aspect remains
  to be understood.
  
\end{itemize}

\acknowledgments{We thank Stephan D\"{u}rr and Christian Hoelbling
for instructive comments.
The code development and testing were performed at the cluster Isabella
of the Zagreb University Computing Centre (SRCE). The production runs
were carried out on the cluster of the Instituto de Ciencias Nucleares,
UNAM. We thank Luciano D\'{\i}az Gonz\'{a}lez for assistance.
This work was supported by the Faculty of Geotechnical Engineering (University
of Zagreb, Croatia) through the project
``Change of the Eigenvalue Distribution at
the Temperature Transition'' (2186-73-13-19-11), and by UNAM-DGAPA
through PAPIIT project IG100219, ``Exploraci\'{o}n te\'{o}rica y
experimental del diagrama de fase de la cromodin\'{a}mica cu\'{a}ntica''.}

\bibliographystyle{JHEP-mcite}
\bibliography{ref}

\end{document}